\documentstyle[preprint,aps,floats,tighten]{revtex}
\input{psfig.tex}
\begin{document}
\draft
\preprint{\vbox{\hbox{USITP-10/96} \hbox{UUITP-1/97} \hbox{CU-TP-810}
    \hbox{CAL-627} \hbox{astro-ph/9702037} }}

\title{Astrophysical-Neutrino Detection with Angular and Energy
  Resolution}

\author{Lars Bergstr\"om\footnote{E-mail address: lbe@physto.se}}
\address{Department of Physics, Stockholm University, Box 6730,
  SE-113~85~Stockholm, Sweden} \author{Joakim Edsj\"o\footnote{E-mail
    address: edsjo@teorfys.uu.se}} \address{Department of Theoretical
  Physics, Uppsala University, Box 803, SE-751 08 Uppsala, Sweden}
\author{Marc Kamionkowski\footnote{E-mail address:
    kamion@phys.columbia.edu}} \address{Department of Physics,
  Columbia University, 538 West 120th Street, New York, New
  York~~10027, U.S.A.}

\date{February 4, 1997}

\maketitle

\begin{abstract}
  We investigate the improvement in sensitivity to astrophysical point
  sources of energetic ($\gtrsim1$ GeV) neutrinos which can be
  achieved with angular and/or energy resolution of the
  neutrino-induced muon.  As a specific example we consider WIMP
  annihilation in the Sun and in the Earth as a neutrino source.
  The sensitivity is improved by using the angular and energy
  distribution to reduce the atmospheric-neutrino background.
  Although the specific improvements depend on the backgrounds and
  assumed sources, the sensitivity to a WIMP signal may be improved,
  with equal exposure, by up to roughly a factor of two with good angular
  resolution, and by up to roughly a factor of three with good energy
  resolution.  In case of a positive detection, energy
  resolution would also improve the measurement of the neutrino
  energy spectrum and therefore provide information on the WIMP
  mass and composition.

\end{abstract}

\long\def\comment#1{} \def\cE{{\cal E}}

\section{INTRODUCTION}

High-energy neutrino astrophysics is a rapidly growing field at the
interface of particle physics and astrophysics.  In addition to the
high-energy neutrinos expected from various cosmic accelerators,
neutrinos with energies in the range of ${\cal O}(1-1000\, {\rm GeV})$
may be produced by the annihilation of weakly-interacting massive
particles (WIMPs) at the core of the Sun and/or Earth if WIMPs
populate the Galactic halo \cite{SOS,jkg}.  Detection of such
neutrinos would be revolutionary for cosmology and for particle
physics.

Several detectors capable of detecting such neutrinos, such as Baksan
\cite{baksan}, IMB \cite{imb}, Kamiokande II \cite{kamiokande}, MACRO
\cite{macro}, and Frejus \cite{frejus} have already put important
constraints on the fluxes of such neutrinos, and the sensitivity
should increase dramatically with the advent of several new
experiments such as Super-Kamiokande \cite{SuperKamiokande}, AMANDA
\cite{amanda}, NESTOR \cite{nestor}, and perhaps others
\cite{wonyong}.

Given a specific WIMP candidate---for example, a neutralino in a
supersymmetric extension of the standard model---calculation of the
differential energy spectrum of the neutrino is straightforward
\cite{RS,joakim}.  Neutrinos are produced by decays of the WIMP
annihilation products, so the detailed energy spectrum depends on the
mass of the neutralino and on its composition.  The mean neutrino
energy generally rises with increasing WIMP mass.

A high-energy neutrino is inferred through observation of an upward
muon produced by a charged-current interaction of the neutrino in the
rock (or ice) below the detector.  (Downward muons can similarly be
produced, but these rare events are usually overwhelmed by the
enormous flux of downward muons from cosmic-ray showers in the
atmosphere.)  Given a point source of neutrinos with some energy
distribution, muons are produced with a distribution of energies and
angles.  The mean muon energy increases with the mean neutrino energy,
and the average angle of production decreases with increasing neutrino
energy.  Unfortunately, although these detectors have decent angular
resolution, they have little or no energy resolution.  The muon range
in matter is roughly proportional to its production energy.  Since we
cannot tell where the muon was produced, we cannot determine its
energy.  All we can say is that if it passes through the entire
detector, it must have an energy greater than some threshold $E_{\rm
  thresh}$ which, for example, is roughly 2 GeV for IMB, Kamiokande
II, and MACRO.

Since the cross section for muon production is proportional to the
energy as is the mean muon range, the probability of detecting a
neutrino depends on the square of the neutrino energy.  Although each
WIMP candidate may produce a different neutrino energy distribution,
the flux of neutrino-induced muons depends only on the second moment
of the neutrino energy spectrum.

If there were no backgrounds, one could simply look for neutrinos by
looking for upward muons from the point source of interest.  Of
course, due to the finite muon production angle, we would want to
accept muons from a large enough solid angle around the point source
to assure that we were getting all the events.  For example, the rms
angle between the neutrino direction and the direction of the induced
muon is $\sim 20^\circ(E_\nu/10\, {\rm GeV})^{-1/2}$.  Furthermore, the
muon typically carries half the neutrino energy, so the angular radius
of the acceptance cone should be $\sim 14^\circ(E_\mu/10\, {\rm
GeV})^{-1/2}$.  A null result after some exposure would then translate
into an upper limit to the second moment of the neutrino energy
spectrum and thus constrain WIMP candidates.

However, neutrino point sources must be distinguished {}from a
background of atmospheric neutrinos with an energy spectrum which
falls roughly as a power law.  The atmospheric-neutrino background is
nearly isotropic (on small enough angular scales), so the flux is
proportional to the solid angle of the acceptance cone around the
point source.  If only the second moment (and no further information)
of the WIMP neutrino energy distribution is specified, the background
flux is that in a cone large enough to include all muons which may
have been produced by a neutrino with energy just large enough to
produce a muon above threshold.  This produces a conservatively large
estimate of the background and thus a conservative sensitivity to the
point source of interest. Improving the sensitivity in this single-bin
approach by optimizing the angular cut has been studied in Ref.~
\cite{BottinoMoscoso} and we will compare with that approach later.

Models for candidate point sources of astrophysical neutrinos (such as
the example of WIMP annihilation in the Sun and Earth on which we
focus here) predict the energy distribution of the neutrinos, and
therefore the muon angular and energy distribution can also be
predicted.  These differ from the angular and energy distribution of
the atmospheric-neutrino background, so experimental determination of
the muon energies and directions can be used to discriminate between
sources and backgrounds.

In this paper we quantify the improvements to the sensitivity of
astrophysical neutrino point sources that can be achieved if the
direction and energy of the neutrino-induced muon can be resolved.  If
the direction of the muons can be measured with higher resolution than
the width of the angular distribution of the signal, the
signal-to-noise ratio can be increased with the same exposure.  This
is analogous to the use of the nuclear recoil-energy distribution in
direct-detection experiments \cite{bottino}.  We also calculate the
further improvement in the sensitivity if the experiment has energy
resolution as well.

We estimate the sensitivity to a signal in several ways.  We will
assume that the background flux is known (we also compare with the
case of it being unknown) and consider the sensitivity to a signal
given either that it comes from a general WIMP or that it comes from a
WIMP with a specific mass and composition.  We find that angular
resolution can improve the sensitivity, for fixed exposure, roughly by
a factor of 2 (depending on backgrounds and signal fluxes), and that
the sensitivity can be improved by up to roughly a factor of 3 if
there is both angular and energy resolution.

In the following Section, we describe how we calculate the expected
sensitivities, and in Section \ref{sec:results} we present numerical
results for some representative models.  We then close with some
concluding remarks in Section \ref{conclusions} and present a table of
numerical results in an Appendix.

\section{Calculational Method}
\label{sec:calcmethod}

\subsection{Introduction}
Consider a theory which predicts an upward-muon flux $\phi_s$ (where
the $s$ stands for ``signal'') with an angular distribution
$d\phi_s/d\theta=\phi_s^0 f_s(\theta) \sin \theta$, where $\theta$ is
the angle the muon makes with the direction of the point source of
interest, and $\int\, f_s(\theta)\,\sin\theta\,d\theta=1$ (i.e.,
$f(\theta)$ is constant for an isotropic distribution).  We would like
to disentangle this signal from a background of atmospheric neutrinos
which has a flux $\phi_b$ with an angular distribution
$d\phi_b/d\theta = \phi_b^0 f_b(\theta) \sin \theta$, which is nearly
isotropic (at least on small angular scales).

Consider first an experiment that can only tell that a muon has been
detected with an angle $\theta\leq\theta_{\rm max}$, but no further
information on the muon direction is available.  Alternatively,
suppose an experiment can only detect muons above some energy
threshold $E_{\rm thresh}$ with no further information on the energy.
If we wish to derive an upper limit to the neutrino flux from a point
source in a model-independent fashion, we must make the most
conservative assumption that all the muon energies were near
threshold.  Then, the angular acceptance cone around the source must
be large enough to include all (or most) of the muons produced by
neutrinos from the point source.  One would therefore have some number
of muons detected with an angle $\theta\leq\theta_{\rm max}$.  For
example, in their searches for energetic neutrinos from the Earth and
Sun, the Baksan collaboration \cite{baksan} reports the flux of muons
within an angle $\theta_{\rm max}=30^\circ$ of the Sun or the center
of the Earth.  The Kamiokande collaboration \cite{kamiokande} reports
the flux of muons within an angle varying between $\theta_{\rm
  max}=5^\circ$--$30^\circ$.

With such a result, the number of background events after an exposure
$\cE$ (for example, in units of km$^2$ yr) is $N_b=\cE \phi_b^0
\int_0^{\theta_{\rm max}}\,f_b(\theta)\,\sin\theta\,d\theta$.  
The number of expected events from the
source of interest is $N_s=\cE\phi_s^0 \int_0^{\theta_{\rm max}}\,
f_s(\theta)\, \sin\theta\, d\theta\simeq \cE\phi_s^0$, where we have
made the reasonable assumption that signal muons come from the
vicinity of $\theta=0$.  A
$3\sigma$ detection would require an excess of $3\sqrt{N_b+N_s}$ events
over the number expected.  Then, a $3\sigma$ excess will be observable
only if $\phi_s > 3\sigma$ where $\sigma = \sqrt{N_b+N_s}/\cE$.  For an
isotropic background, $f_b(\theta)=1/2$, so $\sigma\simeq
[(1-\cos\theta_{\rm max})\phi_b/(2\cE)]^{1/2}$.  Thus, the sensitivity
scales as the inverse square root of the exposure.  Furthermore, since
$\theta_{\rm max}$ scales as $E_{\rm thresh}^{-1/2}$, the sensitivity
of the experiment improves as $E_{\rm thresh}^{-1/2}$, assuming the
energies of most signal muons are above threshold and $\theta_{\rm
  max}\ll1$.

{}From such a simple experiment described above where no energy or
angular distributions are used, we can conclude that the minimal
exposure required for a $3\sigma$ discovery is
\begin{equation} \label{eq:expminsimple}
  \cE_{\rm min} = \frac{9 \left(\phi_b+\phi_s \right)}{\phi_s^2}
\end{equation}
where $\phi_b$ and $\phi_s$ are the background and signal fluxes above
threshold and within the angular cone of acceptance $\theta_{\rm
  max}$.  Note that Eq.~(\ref{eq:expminsimple}) is only valid when the
fluxes are high.  This minimal exposure is relevant to the way, e.g.,
Baksan and Kamiokande have analyzed their data (with different values
of $\theta_{\rm max}$). In the following more detailed examples,
Eq.~(\ref{eq:expminsimple}) with $\theta_{\rm max}=5^\circ$ will be
used for comparison. Note that for low masses ($\lesssim 100$ GeV) the
optimal $\theta_{\rm max}$ will be higher and for high masses it will
be lower. However, one cannot know in advance what the optimal cut
will be. Hence we have chosen 5$^\circ$ as a reasonable compromise
giving decent results both for low and high masses.


\subsection{Covariance-Matrix Analysis}

Now consider a slightly more sophisticated experiment which has
angular and/or energy resolution.  It is possible that we will
actually be fitting for both a background and a signal flux of muons
where the background flux is given by
\begin{equation}\label{eq:eaatmbg}
     \frac{d^2 \phi_b}{dE d\theta}(E,\theta) = \phi_b^0 f_b(E,\theta),
\end{equation}
which we assume to be isotropic (at least over small angular patches),
$f_b(E,\theta)=f_b(E)$.
We will only consider the atmospheric background resulting from
cosmic-ray interactions in the Earth's atmosphere. This is an
irreducible background that cannot be avoided even in very deep
underground detectors.  In addition, we will want to fit data for an
annihilation signal which generally depends on the WIMP mass $m_\chi$
and on the WIMP composition.  We may parameterize this as
\begin{equation} \label{eq:param}
     \frac{d^2 \phi_s}{dE d\theta} (E,\theta) = \phi_s^0
     \left[ a f_{\rm hard}(m_\chi,E,\theta)+(1-a) 
     f_{\rm soft}(m_\chi,E,\theta) \right] ,
\end{equation}
where $a$ parameterizes the relative contributions of a ``hard'' and
``soft'' annihilation spectrum. As a ``hard'' annihilation spectrum we
have used the $\tau^+ \tau^-$ channel below the $W$-mass and $W^+ W^-$
above and as a soft spectrum we have used $b \bar{b}$.  These channels
represent the extreme hardnesses of the spectrum for any given WIMP
mass. For the evaluation of the neutrino and muon flux for these
channels we have used the method given in Ref.\ \cite{joakim}, where
the whole chain of processes from the annihilation products in the
core of the Sun or the Earth to detectable muons at the Earth was
considered.

Therefore, we are assuming that the muon angular and/or energy
distribution from both background and signal will be described by the
set of parameters, ${\bf s}= \{\phi_b^0,\phi_s^0,m_\chi,a\}$ (one
could also envision more parameters).  We now want to ask, with what
precision can we measure these parameters with a given experiment,
assuming the true distribution is given by some set of parameters,
${\bf s}_0$?

To do so, we assume the data is binned into a number of angle/energy
bins, and each bin $i$ is centered on angle $\theta_i$ and energy
$E_i$ with widths $\Delta E_i$ and $\Delta\theta_i$.  Therefore, for a
given set ${\bf s}$ of parameters, the flux will be
\begin{equation}
  \frac{d^2 \phi}{dE d\theta} (E,\theta;{\bf s}) = \frac{d^2\phi_b}{dE
    d\theta} (E,\theta;{\bf s})+\frac{d^2 \phi_s}{dE d\theta}
    (E,\theta;{\bf s}),
\end{equation}
where we
have written the dependence of the flux on the model parameters ${\bf
  s}$.  The probability distribution for the number of events expected
in each bin is a Poisson distribution with mean $N_i = {\cal E}
\frac{d^2 \phi}{dE d\theta} (E_i,\theta_i) \Delta E_i\Delta \theta_i
$, so it has a width $\sigma_i =\sqrt{N_i}$.

So, suppose the true parameters are ${\bf s}_0$.  Then the probability
distribution for observing an angle/energy distribution which is best
fit by the parameters ${\bf s}$ is
\begin{equation}
     P({\bf s}) \propto \exp \left[ -{1\over2} ({\bf s} - {\bf
     s}_0) \cdot [ \alpha ] \cdot ({\bf s} - {\bf s}_0) \right],
\end{equation}
where the curvature matrix $[\alpha]$ is given approximately by
\begin{eqnarray}
     \alpha_{ab} & = & {\cal E}\sum_i\, {1\over \sigma_i^2} \, {\partial N_i
     \over \partial s_a}\, {\partial N_i \over
     \partial s_b}   \nonumber \\
     & = &  4{\cal E}\sum_i\, {\partial \sqrt{N_i}
     \over \partial s_a}({\bf s}_0) \, {\partial \sqrt{N_i} \over \partial
     s_b}({\bf s}_0),
\end{eqnarray}
where the partial derivatives are evaluated at ${\bf s}= {\bf s}_0$,
and we used $\sigma_i = \sqrt{N_i}$ in the second line.  In a
realistic experiment, the width of the bins would be comparable to the
angular and/or energy resolution of the experiment.  In the limit of
perfect angular and energy resolution, the sum becomes an integral,
\begin{equation}
     \alpha_{ab}  = 4{\cal E} \int\int\, dE\, d\theta \, {\partial
     \sqrt{d^2\phi(E,\theta;{\bf s})/dE d\theta} \over \partial s_a}\,
     {\partial \sqrt{d^2 \phi(E,\theta;{\bf s})/dE d\theta} \over \partial
     s_b}.
\label{integralalpha}
\end{equation}
The covariance matrix, $[{\cal C}] = [\alpha]^{-1}$ gives an estimate
of the standard errors that would be obtained from a
maximum-likelihood fit to data: The standard error in measuring the
parameter $s_a$ (after marginalizing over all the other undetermined
parameters) is approximately $\sigma_a \simeq {\cal C}_{aa}^{1/2}$.
If three times the standard error in $\phi_s^0$ is less than
$\phi_s^0$, for a given underlying model ${\bf s}_0$ and for a given
experiment, then this model will be distinguishable from background at
the $3\sigma$ level.

If all of the parameters except for $\phi_s^0$ are fixed, then
$[\alpha]$ is a $1\times1$ matrix, i.e., $1/\sigma^2$.  In this
case, Eq. (\ref{integralalpha}) reduces to
\begin{equation}
    {1\over \sigma^2} = \cE \int\int\, {
    [f_s(\theta,E)]^2 \over \phi_b^0 f_b(E) + \phi_s^0
    f_s (\theta,E)}
    \, \sin\theta\, d\theta\, dE.
\label{energysensitivitytwo}
\end{equation}
To illustrate, if there were no background, Eq.
(\ref{energysensitivitytwo}) says that the statistical uncertainty in
the number of events is the square root of the number of events, and
this makes sense.  However, if the total number of events is nonzero,
then a signal has been discovered.  In other words, a 99\% CL does not
correspond to $3\sigma$ for small numbers.

In fact, if there is no background, and an event is seen, it
constitutes discovery.  On the other hand, if nothing is seen, the
95\% CL upper limit to the number is 3.


\section{Results}
\label{sec:results}

We are now ready to perform some actual calculations using the
techniques described in the previous Section for the specific example
of WIMP annihilation in the Sun and Earth. WIMPs which are
gravitationally trapped in the Sun/Earth can annihilate to produce
neutrinos which reach a neutrino telescope, interact, and form
detectable muons.  For the muon fluxes we have used the method given
in Ref.\ \cite{joakim} where all relevant processes from the WIMP
annihilation products to the resulting muon flux at a detector are
calculated using Monte Carlo simulations. The muon fluxes are
calculated for different WIMP masses and annihilation channels. We
assume that the neutrino energy spectra are either the hard or soft
spectra described above; energy spectra from realistic WIMP candidates
should fall somewhere between these two extremes. Since the muon flux
is proportional to the neutrino energy squared, the hard annihilation
channels will generally be more important and hence in general the
muon spectra will be more hard than soft. Because of the steep fall
with energy of the atmospheric background, hard spectra generally
require less exposure.  In all integrations with angular (and energy)
distribution the integration in Eq.~(\ref{integralalpha}) is performed
up to $\theta=30^\circ$.
For the atmospheric background we have used the results given in Ref.\ 
\cite{honda}. Even though the absolute value of the background flux is
uncertain by some 20\%, the overall level will certainly be
measured with high accuracy by the new experiments.

In Tables~\ref{tab:suver}--\ref{tab:eaver} in the Appendix, the
minimal exposures required for a $3\sigma$ discovery are given for
detectors with no angular resolution, only angular resolution and
angular and energy resolution assuming that all four parameters in
Eq.~(\ref{eq:param}) are unknown, only the three signal parameters are
unknown and only $\phi_s^0$ is unknown.  In
Figs.~\ref{fig:angular}--\ref{fig:energy} some illustrative examples
of these results are shown and in the following subsections these
results are described and discussed.

\subsection{Detector with angular resolution}
\label{ss:angres}

\begin{figure}
  \centerline{\psfig{file=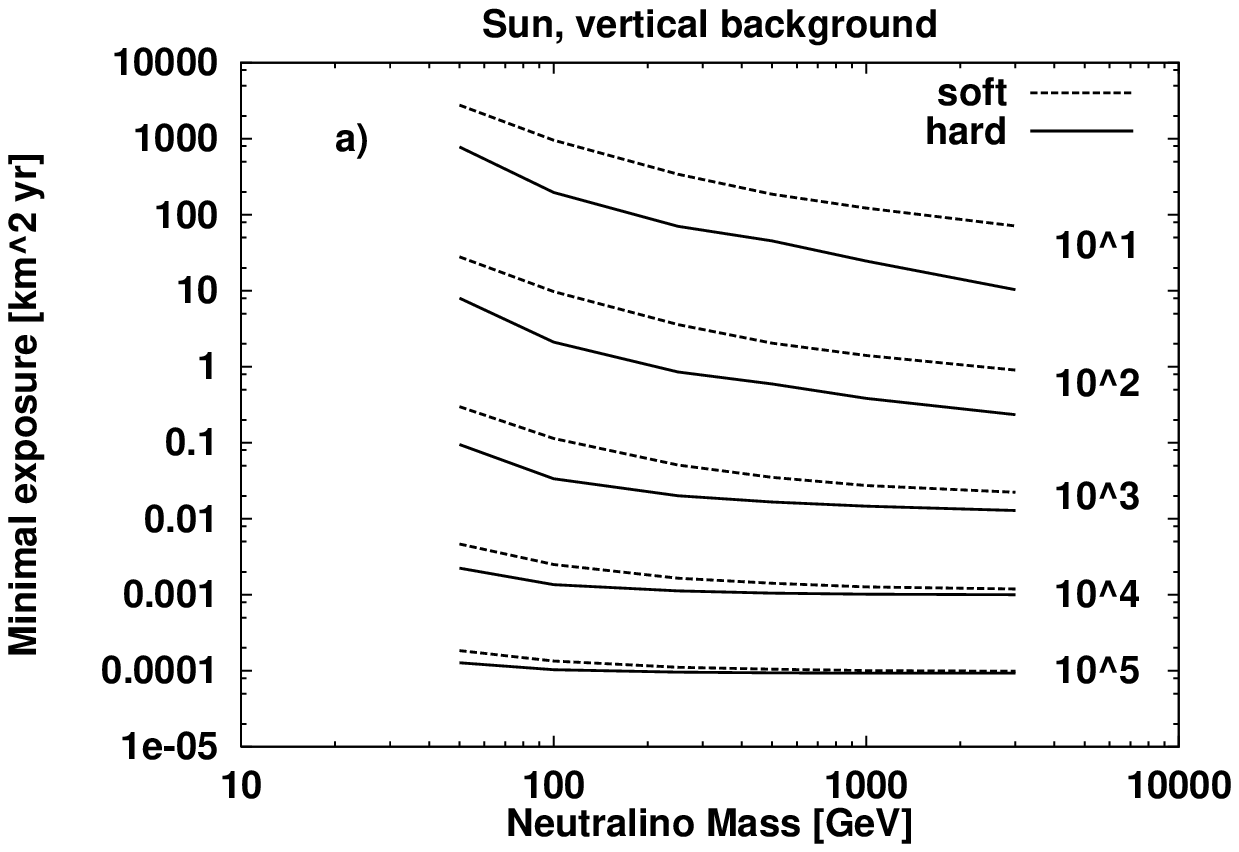,width=3.3in}} \smallskip
  \centerline{\psfig{file=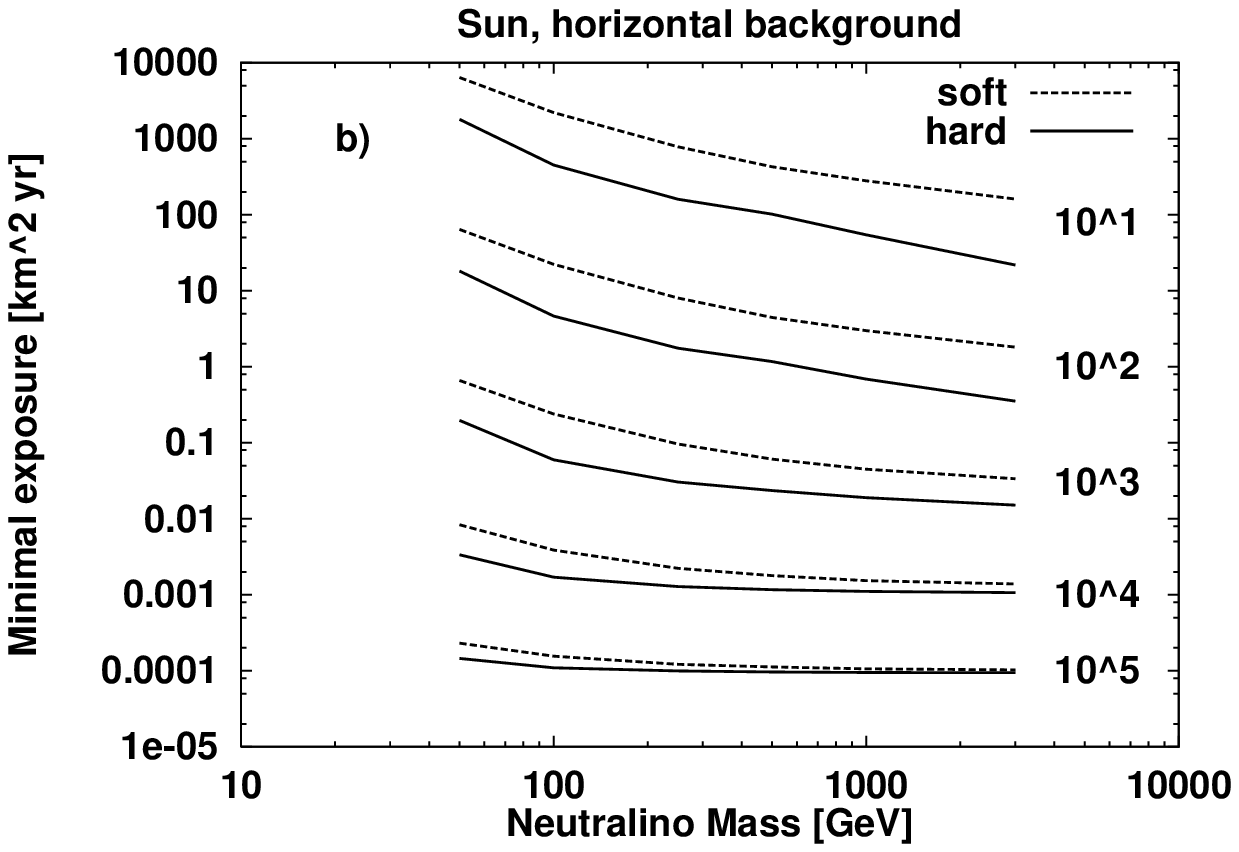,width=3.3in}} \smallskip
  \centerline{\psfig{file=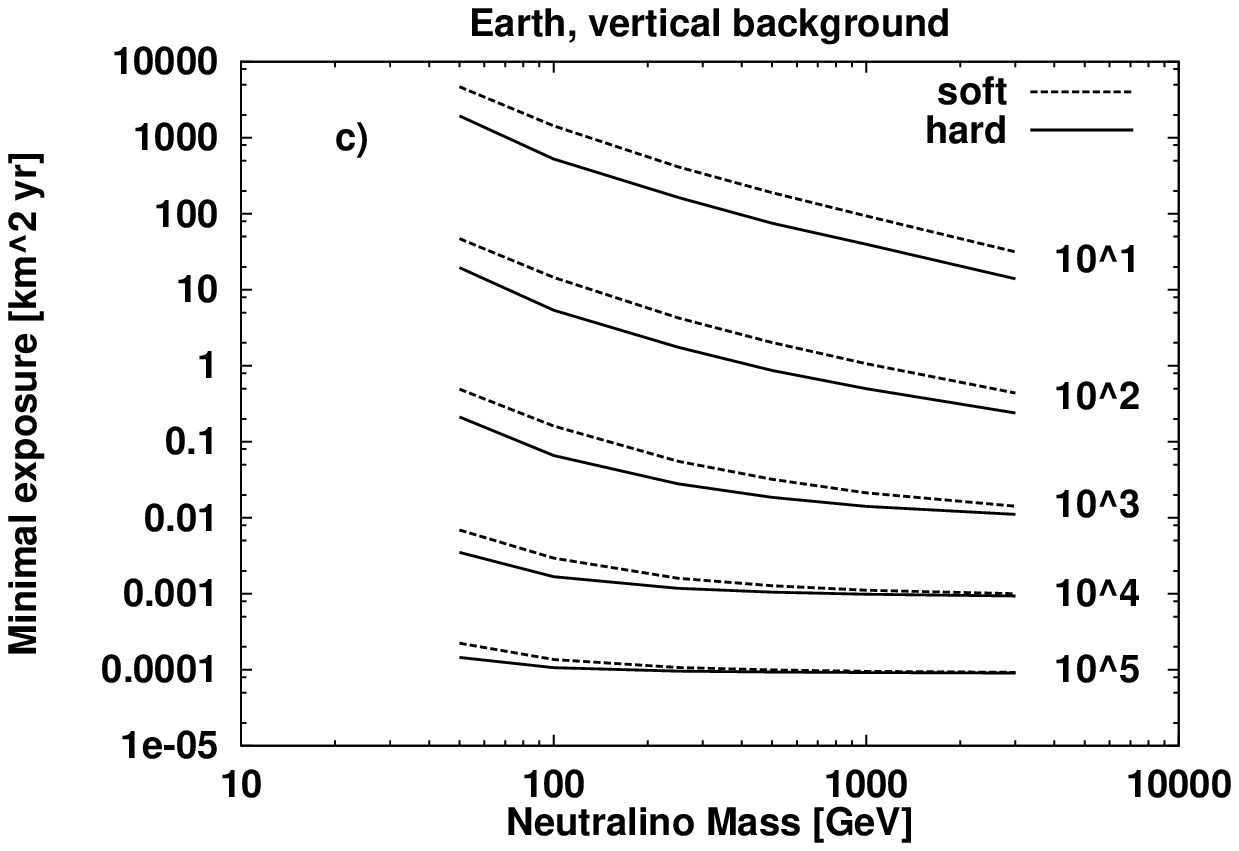,width=3.3in}} \bigskip
\caption{
  The exposures needed for a 3$\sigma$ discovery for different signal
  fluxes (indicated to the right in the figure in units of km$^{-2}$
  yr$^{-1}$) as a function of WIMP mass assuming perfect angular
  resolution but no energy resolution (and with a muon energy
  threshold of 1 GeV). The three figures correspond to annihilation in
  a) the Sun with vertical background, b) the Sun with horizontal
  background and c) the Earth with vertical background.  The
  solid(dashed) lines correspond to soft(hard) muon spectra.  The
  three signal parameters $\{\phi_s^0,m_\chi,a\}$ in
  Eq.~(\protect\ref{eq:param}) are assumed to be unknown while the
  background flux is assumed to be known.  Note that only exposures
  less than, say, 25 km$^{-2}$ yr$^{-1}$ are realistic in the near
  future.  }
\label{fig:angular}
\end{figure}

\begin{figure}
  \centerline{\psfig{file=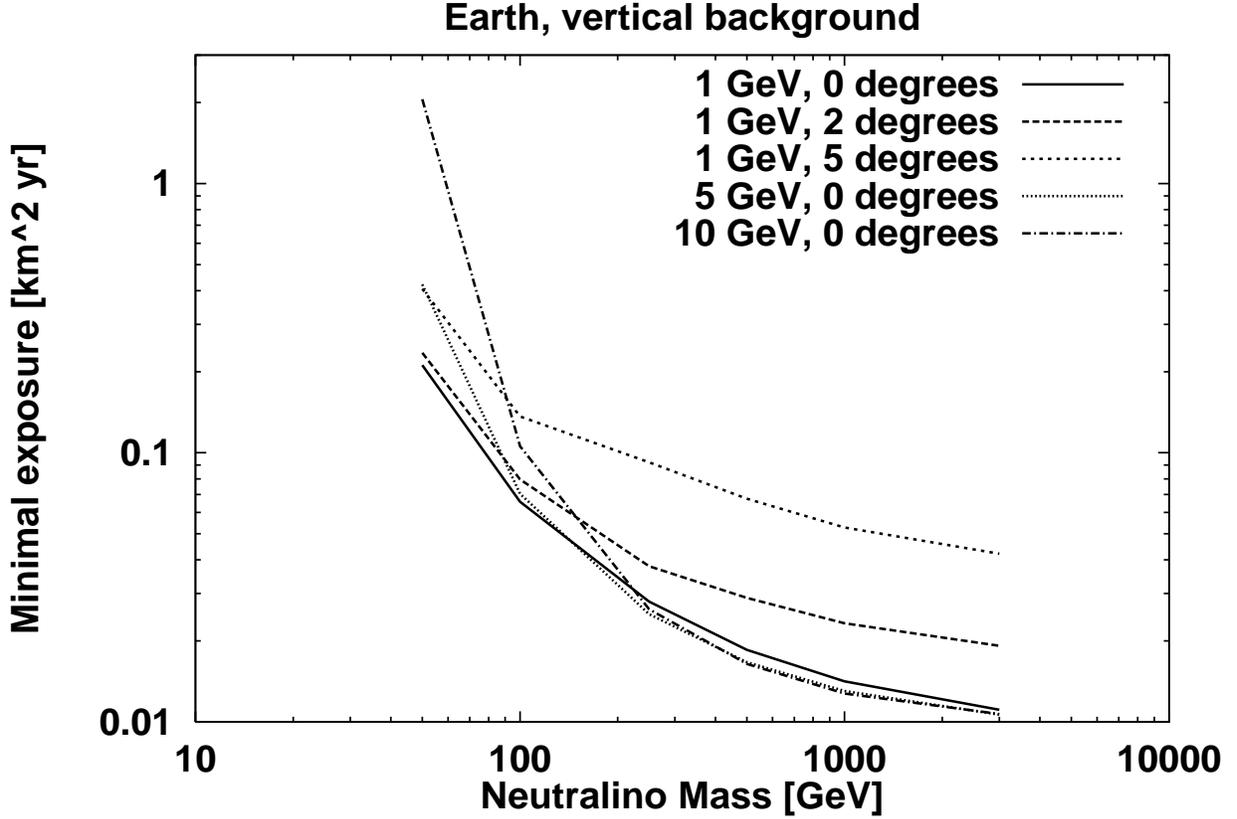,width=\textwidth}} \bigskip
\caption{
  The exposures needed for a 3$\sigma$ discovery for the signal flux
  $\phi_s^0 = 10^3$ km$^{-2}$ yr$^{-1}$ coming from WIMP annihilation
  in the Earth.  The three signal parameters $\{\phi_s^0,m_\chi,a\}$
  in Eq.~(\protect\ref{eq:param}) are assumed to be unknown while the
  background flux is assumed to be known. The change of the minimal
  exposures needed when changing the experimental angular resolution
  or increasing the energy threshold is indicated. All curves are for
  hard annihilation spectra.}
\label{fig:angularethsmear}
\end{figure}

\begin{figure}
  \centerline{\psfig{file=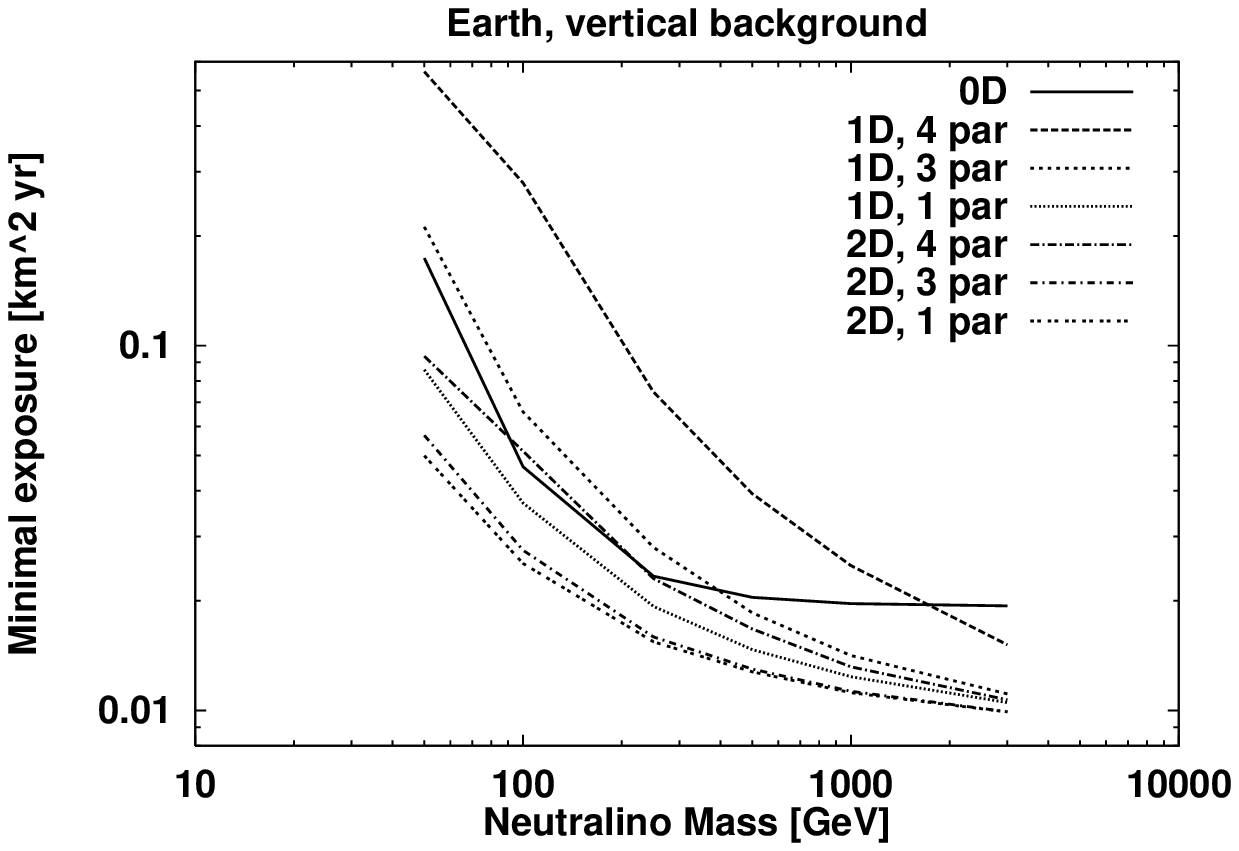,width=\textwidth}} \bigskip
\caption{
  The exposures needed for a 3$\sigma$ discovery for the signal flux
  $\phi_s^0 = 10^3$ km$^{-2}$ yr$^{-1}$ coming from WIMP annihilation
  in the Earth.  The minimal exposures needed for a detector with
  neither angular nor energy resolution (0D), only angular but no
  energy resolution (1D) and both angular and energy resolution (2D)
  is shown. For the 1D and 2D cases, results are given for all four
  parameters in Eqs.~(\protect\ref{eq:eaatmbg}) and
  (\protect\ref{eq:param}) being free (4 par), only the three signal
  flux parameters being free (3 par) and only the normalization of the
  signal flux, $\phi_s^0$ being free (1 par). An energy threshold of 1
  GeV is used in all cases and for the 0D case an integration of the
  fluxes up to $\theta_{\rm max}=5^\circ$ is performed. All curves are
  for hard annihilation spectra.}
\label{fig:energy}
\end{figure}

If we assume that the background flux is known (by, e.g., an
off-source measurement) and that all three parameters in
Eq.~(\ref{eq:param}) corresponding to the signal flux are unknown we
need at least the exposures given in Fig.~\ref{fig:angular} to be able
to make a 3$\sigma$ discovery if we have perfect angular resolution
but no energy resolution.

Note that the spread between soft and hard spectra is due to our
ignorance of the actual branching ratios into different annihilation
channels and the two curves for soft and hard should thus be treated
as extreme values. However, as explained earlier, in general spectra
will be closer to being hard than being soft. We also see that when
the signal flux is high, the difference between soft and hard spectra
does not matter since in this case the signal-to-noise ratio is high
even though the spectra are smeared.

By comparing Figs.~\ref{fig:angular}(a) and (b), we see that the
exposures needed when looking at the Sun horizontally are about a
factor of 2 higher when the signal flux is low due to the atmospheric
background being higher in the horizontal direction than in the
vertical. When the signal flux is high, the needed exposures are about
the same.

If we compare Figs.~\ref{fig:angular}(a) and (b) with (c), we find
that the curves for the Earth are more dependent on the mass than the
curves for the Sun, being higher at low masses and lower at high
masses. At low masses this difference is because the size of the
annihilation region is non-negligible in the Earth and is thus making
the angular distributions wider. For high masses the difference is due
to the fact that neutrino interactions on the way out of the Earth are
negligible while they are not for the Sun thus softening the Sun
spectra at high masses.

So far we have considered a detector with perfect angular resolution.
In Fig.~\ref{fig:angularethsmear} we show an example of what happens
if we add experimental angular resolution. As expected the needed
exposures are higher at high masses since we now cannot make use of
the highly collimated signal flux at these high masses. The general
trend for other fluxes and for the Sun is the same as that shown in
the Figure.  In the Figure, curves for different energy thresholds are
also given. These are described in next subsection.

In Fig.~\ref{fig:energy} we evaluate the sensitivities attainable
if we marginalize over $m_\chi$ and $a$ (``3 par'') and if we
marginalize over $\phi_b^0$ as well (``4 par'').  To compare, we
also show the sensitivities assuming these parameters are fixed
and we are just fitting for the source flux $\phi_s^0$ (``1
par'').  We also compare with the case of just using
one bin up to $\theta_{\rm max}=5^\circ$ and checking if any signal
above background can be seen in this single bin. We see that we gain a
factor of 2--3 by knowing the background flux in advance (which is
quite reasonable).  Note that the position of the 4-parameter curves
all depend on the upper limit of the $\theta$-integration since the
higher it is, the more background is included in the fit and the lower
the curves get.  If we compare the improvement by using the
3-parameter fit for the signal to the simple case of using just one
bin up to a certain angle $\theta_{\rm max}$ we see that the minimal
exposures needed are of the same order for small masses while there is
an improvement of up to a factor of 2 at higher masses. Note however
that with the 3-parameter approach we will also gain some information
on the mass and hardness of the spectrum. Also, a problem with the
single bin approach is the choice of $\theta_{\rm max}$. If we choose
$\theta_{\rm max}<5^\circ$ we could get the single bin curve and the
3-parameter angular resolution curve to match at high masses but then
the single bin curve would do much worse at low masses.  At low masses
($\lesssim100$ GeV) the optimum choice would be more than $5^\circ$
and at high masses less. We cannot however know in beforehand what the
optimum choice is.  This problem we avoid by using the signal flux
parameterization, Eq~(\ref{eq:param}), proposed here and get about the
same sensitivity as with the single bin approach with optimal
$\theta_{\rm max}$.

If we are interested only in looking for a nonzero signal flux
$\phi_s^0$ for some specific hypothesized WIMP mass and neutrino
spectrum, we gain about a factor of 1--1.5 more and will always do
better than the one-bin approach. This might be the case if the WIMP
is found at accelerators and we know its properties but want to know
if it constitutes the dark matter in the Universe.

We can conclude from Fig.~\ref{fig:angular} and
Tables~\ref{tab:suver}--\ref{tab:eaver} that a reasonable detector
with perfect angular resolution and exposures in the order of 1--25
km$^2$ yr would be able to detect signal fluxes down to about 100
km$^{-2}$ yr$^{-1}$ (slightly more at low masses and slightly less at
high masses).

\subsection{Detector with angular resolution and variable energy 
  threshold}

Our next case to consider is a neutrino telescope with perfect angular
resolution and variable muon energy thresholds.  This may be achieved
by, e.g., using different triggering conditions. By increasing the
energy threshold we expect to increase the signal-to-noise ratio (at
least if the threshold is well below the WIMP mass) and hence get
better signal detection possibilities. Of course, it would be even
better to have full energy resolution (as described in the next
subsection).  However, the energy resolution of current neutrino
detectors is not great, but changing the threshold might be a good
option.

We have chosen to evaluate the needed minimal exposures for the muon
energy thresholds $E_{\mu}^{th}$ = 1, 5, 10, 25, 50, and 100 GeV and in
Fig.~\ref{fig:angularethsmear} we show the minimal exposures needed
for detection for muon energy thresholds of 5 and 10 GeV compared to
the threshold of 1 GeV for WIMP annihilation in the Earth. The general
trend for annihilation in the Sun and for other signal fluxes are the
same as those shown in the figure, namely that there is a small gain
by increasing the threshold when the WIMP mass is above 100 GeV but
below too much signal is also lost and the needed minimal exposures
are higher. At thresholds of 25, 50, and 100 GeV, the needed minimal
exposures are higher than or about the same as with a threshold of 10
GeV. Hence there is no gain with increasing the threshold higher than
to about 10 GeV.

To conclude on varying the threshold, the gain is very small and only
for WIMP masses above 100 GeV. Increasing the threshold above 10 GeV
gives no further improvements. On the other hand, large detectors
(like AMANDA) which have a threshold of tens of GeV will not lose much
sensitivity either, for WIMP masses above 100 GeV.

\subsection{Detector with angular and energy resolution}

If we imagine a detector where we have both perfect angular and energy
resolution, what improvement in minimal exposures do we get?  In
Fig.~\ref{fig:energy} we compare the minimal exposures needed for such
a detector and a detector with only angular resolution. We also
compare with the simple case of just having one bin in angle. The
improvement by having both energy and angular resolution compared to
just having angular resolution can be as high as a factor of 2 at low
WIMP masses, but at higher masses the improvement is less. However, if
the signal flux is small the improvement is of about a factor of 2 at
all masses and if the signal flux is high, the improvement is small
for all masses.

Since supersymmetric models generally predict relatively low fluxes
\cite{SOS,jkg}, this improvement could be significant.

\subsection{Example of a neutrino detector}

\begin{figure}
  \centerline{\psfig{file=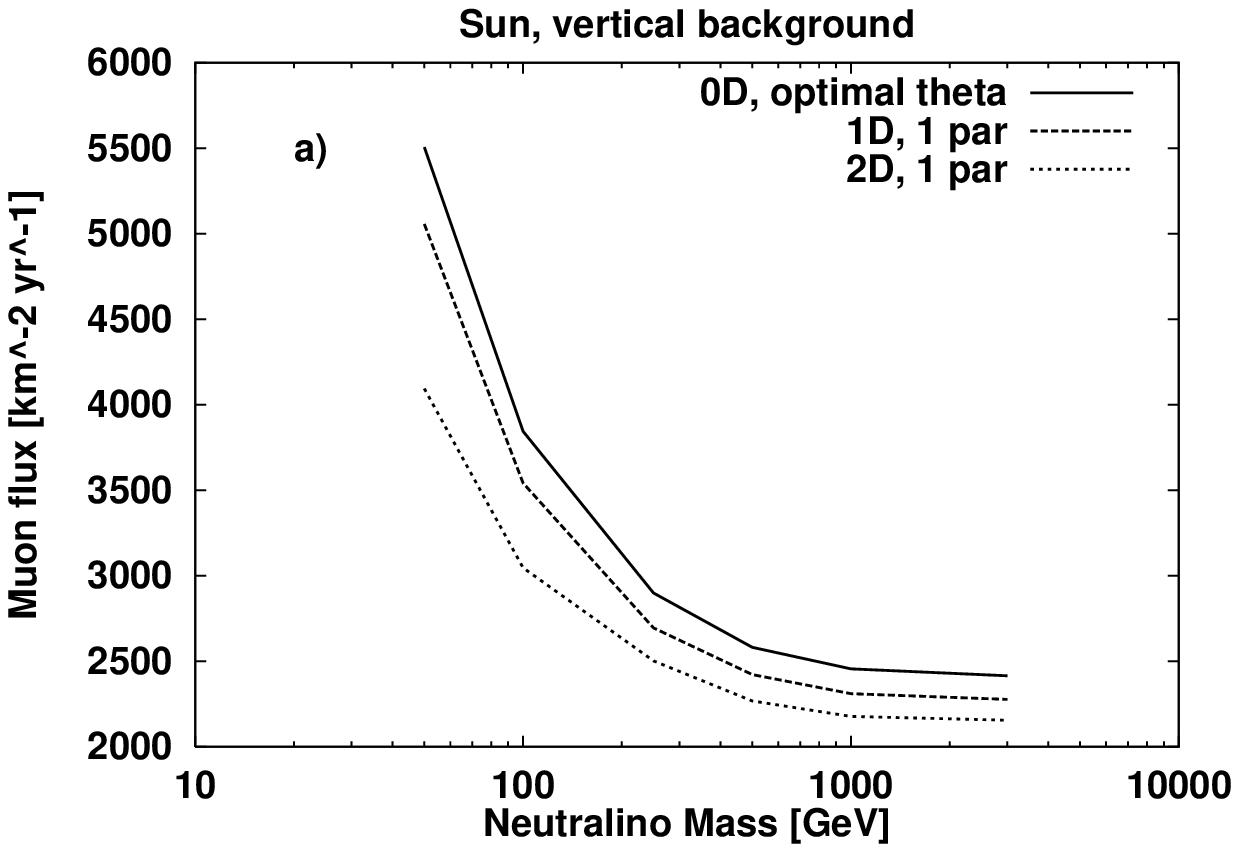,width=0.75\textwidth}} \bigskip
  \centerline{\psfig{file=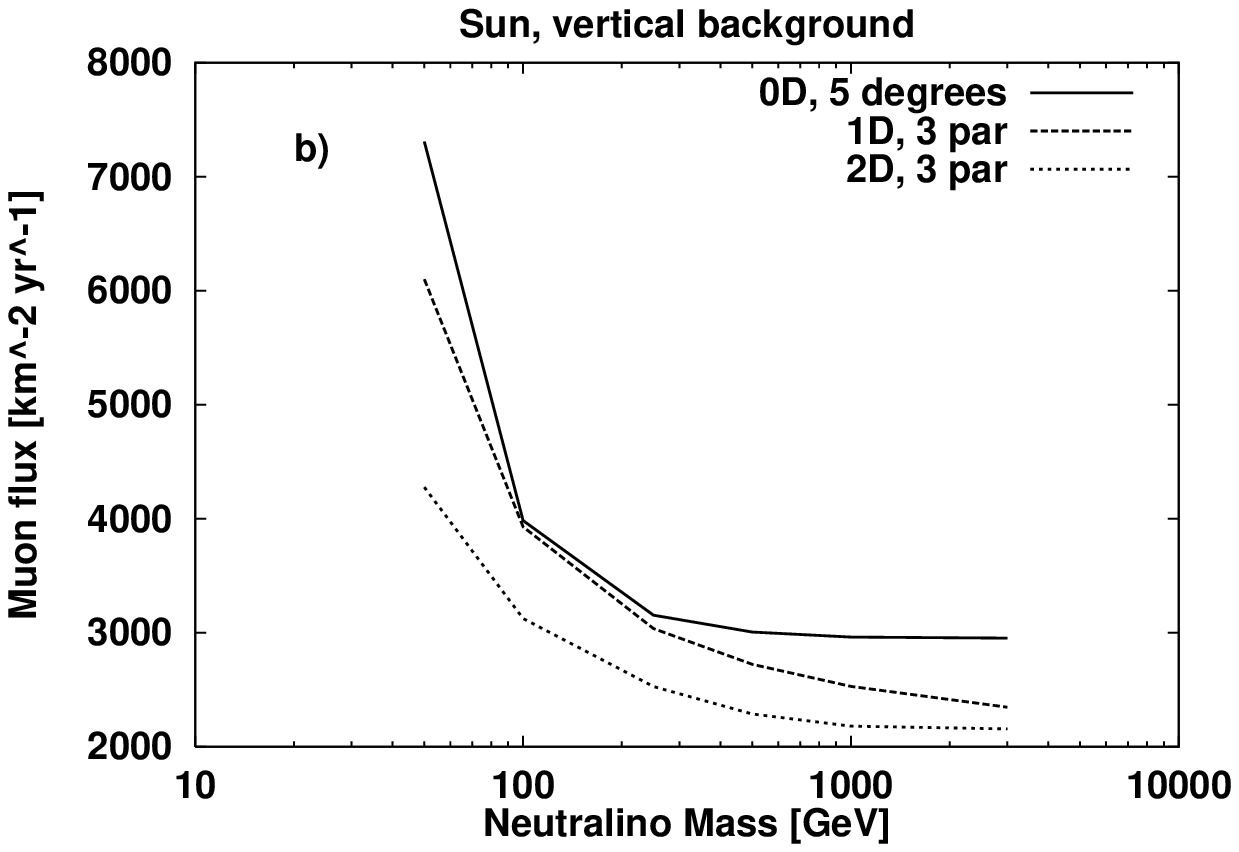,width=0.75\textwidth}} \bigskip
\caption{
  The muon fluxes (coming from WIMP annihilations in the Sun) that can
  be discovered (at the 3$\sigma$ level) when the WIMP spectrum is
  hard with the exposure $\cE = 0.005$ km$^{2}$ yr. The different
  cases of only having one angular bin (0D), having perfect angular
  resolution (1D) and having perfect angular and energy resolution
  (2D) are compared.  In a) the optimal $\theta_{\rm max}$ for each
  mass is used in the 0D case and only $\phi_s^0$ is assumed to be
  unknown for the 1D and 2D cases and in b) $\theta_{\rm max}=5^\circ$
  is used in the 0D case and all three signal parameters
  $\{\phi_s^0,m_\chi,a\}$ in Eq.~(\protect\ref{eq:param}) are assumed
  to be unknown in the 1D and 2D cases.  In both a) and b) the
  background flux is assumed to be known and the atmospheric
  background is assumed to be vertical.}
\label{fig:macro}
\end{figure}

Let us consider a detector with a size of MACRO (effective area about
650 m$^2$ towards the Sun). What improvements can be done with such a
detector by using only angular or both the angular and energy
resolution. Assume that the exposure towards the Sun is $\cE=5000$
km$^2$ yr (about 15 years of data taking and the Sun being below the
horizon for 50\% of the time) and that the atmospheric background is
vertical. In Fig.~\ref{fig:macro} we show a comparison between having
only one bin, having perfect angular resolution and having both
perfect angular and energy resolution.  The signal fluxes that are
given here are for hard muon fluxes. When the fluxes are soft, the
curves are higher. In (a), we show results for the case when we have a
given hypothesis, i.e.\ a WIMP of a given mass and a given
composition. Hence we compare the one-bin approach with optimal
$\theta_{\rm max}$ (for the given hypothesis) with the case of angular
and/or energy resolution where only $\phi_s^0$ is kept free.  In (b),
we show results when our hypothesis is that we look for WIMPs with
masses up to 3000 GeV and with different composition. Hence we compare
results for $\theta_{\rm max}=5^\circ$ and all three signal parameters
$\{\phi_s^0,m_\chi,a\}$ in Eq.~(\protect\ref{eq:param}) being unknown.
We see clearly that at high masses, there is a substantial gain by
having perfect angular resolution (at what masses the gain is highest
depends on $\theta_{\rm max}$) and especially at low masses there is a
substantial gain by having perfect energy resolution as well.

Note that in this example we are almost in the signal dominated regime
so that by increasing the exposure, $\cE$, by a small factor, we
decrease the limit on $\phi_s^0$ by almost the same amount,
especially at high masses.

\subsection{Discussion}

{}From our results earlier in this Section we see that for the
neutrino signal coming from WIMP annihilations in the Sun and in the
Earth we gain about a factor of 2 at high masses by using the angular
resolution of a neutrino telescope. At lower masses there is hardly
any difference to the simple approach of just having one bin up to a
certain maximum angle $\theta_{\rm max}$, but at higher masses there
is a significant difference.  Note however that how well the
single-bin approach does at different masses depends on $\theta_{\rm
  max}$ as described in Section \ref{ss:angres}.  Note also that by
using the parameterization of the signal flux proposed here we can
also gain some information on the WIMP mass and the hardness of the
spectrum.

By varying the energy threshold not much more is gained, but by having
energy resolution about a factor of 1.5--2 can be gained, slightly
more at low signal fluxes and slightly less at high signal fluxes.

We can also note that for neutrino telescopes in the size of about 1
km$^2$ the signal fluxes we can expect to probe is in the region of
50--100 km$^{-2}$ yr$^{-1}$ with a neutrino telescope with angular
resolution and slightly less for a telescope with also energy
resolution.  Hence the signal fluxes within reach are almost an order
of magnitude larger than the expected background coming from cosmic
ray interactions in the Sun (about 15 events km$^{-2}$ yr$^{-1}$
\cite{Sunbgd}). Therefore it is quite safe to neglect this background
at the present stage. When detectors are getting even bigger it will
however be a severe limitation when looking for the neutrino flux from
the Sun since this background is also highly directional as the
signal. Note, though, that the energy dependence is expected to be
quite different, so having energy resolution may be quite beneficial
in this case.

\section{Conclusions}
\label{conclusions}

We have evaluated the improvement in sensitivity to astrophysical
neutrino point sources that can be achieved with muon resolution and
with muon energy resolution.  We have focused on neutrinos from WIMP
annihilation in the Sun and Earth and considered WIMP candidates with
a variety of masses and neutrino spectra.  For example, for detectors
with exposures comparable to Baksan, Kamiokande II, and/or MACRO, an
analysis which uses angular information can improve the sensitivity to
neutrinos from annihilation of WIMPs in the Sun and Earth by
10--40\%, depending on whether the backgrounds are vertical
or horizontal.  Energy resolution could improve the sensitivity by
roughly another 10--65\%.  Angular and energy resolution generally
provides an improvement in sensitivity to
signals which are small (approaching the inverse exposure of the
detector).  We have shown that with a 1 km$^2$ neutrino telescope it
is possible to probe muon fluxes coming from WIMP annihilation in the
Sun and Earth down to about 50--100 km$^{-2}$ yr$^{-1}$.  For
detector with an exposure comparable to that of MACRO after 15
years, energy resolution could in some cases reduce the
atmospheric-neutrino background to insignificant levels.

Currently, the primary issue in dark-matter detection is discovery, so
we have focused here on how the sensitivity to neutrinos from WIMP
annihilation can be improved from energy and angular resolution.
However, directional and energy information will also be useful for
other reasons as well.  For example, in case of a positive detection,
the angular distribution of the muons from the Sun/Earth can be used
to infer the mass of the WIMP \cite{paolo}.  If there is also energy
resolution, this mass determination can be made increasingly precise,
and one might also be able to infer something about the neutrino
energy distribution.  Although we have not carried out the
calculation, the covariance-matrix formalism described in Section
\ref{sec:calcmethod} can be used to evaluate the accuracies with which
the WIMP mass can be measured for various WIMP candidates and with
various experimental configurations.  Of course, the method developed
here can be used the determine the sensitivity to other point sources
of neutrinos as well.  Energy resolution will also be essential for
studying, for example, more conventional astrophysical point sources
of neutrinos and atmospheric neutrinos.

The angular resolution of current experiments is already quite good,
and this information should be used to improve the sensitivity to
neutrinos from dark-matter annihilation.  Although current experiments
do not have good energy resolution (or any energy resolution at all),
there are indeed ideas for obtaining some estimates of the energy, and
ideas for future experiments with fairly precise energy resolution are
currently being explored.  We hope that the work presented here will
help spur new ideas for experimental determination of the muon energy
in neutrino telescopes.

\acknowledgments

LB was supported by the Swedish Natural Science Research Council
(NFR).  MK was supported in part by the D.O.E. under contract
DEFG02-92-ER 40699, NASA under contract NAG5-3091, and by the Alfred
P.\ Sloan Foundation.  MK acknowledges the hospitality of the Theory
Division at CERN, where part of this work was completed.  MK also
acknowledges the hospitality of the Uppsala/NORDITA Astroparticle
Workshop where this work was initiated. MK thanks
Wonyong Lee for useful discussions.

\clearpage
\appendix

\section{Tables of minimal exposures}
\label{app:tables}

Below are tables giving minimal exposures needed for a $3\sigma$
discovery of the neutrino signal coming from WIMP annihilation in the
Sun and Earth. The column labeled ``OD'' refers to an experiment with
a 1-GeV muon threshold but with no further energy or angular
resolution.  The columns labeled ``1D'' refer to experiments with
angular but no energy resolution, and those labeled ``2D'' refer to
experiments with both energy and angular resolution.  The column
labeled ``4par'' gives the minimum exposure needed for a $3\sigma$
detection of the source flux after marginalizing over the unknown WIMP
mass ($m_\chi$), background flux ($\phi_s^0$), and ``hardness'' ($a$)
of the source spectrum.  The column labeled ``3par'' gives the minimum
exposure needed for a $3\sigma$ detection of the source flux assuming
the background flux is known from, e.g., off-source measurements, but
after marginalizing over the unknown WIMP mass and ``hardness'' of the
source spectrum.  The column labeled ``1par'' gives the minimum
exposure needed for a $3\sigma$ detection of the source flux for a
fixed WIMP mass, source-spectrum hardness, and background flux.

Note that if interpolations in the tables are needed, they should be
done on the logarithms of $\phi_s^0$ and $\cE_{\rm min}$.  For smaller
signal fluxes than those given here, $\cE_{\rm min}$ scale such that
lowering the signal flux a factor of 10 increases $\cE_{\rm min}$ by a
factor of 100.

\begin{table}[h]
\begin{tabular}{lrllllll}
  \multicolumn{1}{l}{$\phi_s^0$} & 
  \multicolumn{1}{l}{$m_{\chi}$} & 
  \multicolumn{1}{l}{$\cE_{\rm min}^{\rm 0D}$} & 
  \multicolumn{1}{l}{$\cE_{\rm min}^{\rm 1D, 4 par}$} & 
  \multicolumn{1}{l}{$\cE_{\rm min}^{\rm 1D, 3 par}$} & 
  \multicolumn{1}{l}{$\cE_{\rm min}^{\rm 1D, 1 par}$} &
  \multicolumn{1}{l}{$\cE_{\rm min}^{\rm 2D, 3 par}$} &
  \multicolumn{1}{l}{$\cE_{\rm min}^{\rm 2D, 1 par}$} \\
  \multicolumn{1}{l}{[km$^{-2}$ yr$^{-1}$]} &
  \multicolumn{1}{l}{[GeV]} &
  \multicolumn{1}{l}{[km$^{2}$ yr]} &
  \multicolumn{1}{l}{[km$^{2}$ yr]} &
  \multicolumn{1}{l}{[km$^{2}$ yr]} &
  \multicolumn{1}{l}{[km$^{2}$ yr]} &
  \multicolumn{1}{l}{[km$^{2}$ yr]} &
  \multicolumn{1}{l}{[km$^{2}$ yr]} \\ \hline
  1.0$\times 10^0$ &   50 & 7.3$\times10^4$ &
  3.1$\times 10^5$ & 7.8$\times 10^{4}$ & 
  4.8$\times 10^4$ & 2.1$\times 10^4$ & 1.9$\times 10^4$ \\
  1.0$\times 10^0$ &  100 & 2.0$\times10^4$ &
  1.0$\times 10^5$ & 2.0$\times 10^{4}$ & 
  1.4$\times 10^4$ & 6.1$\times 10^3$ & 5.4$\times 10^3$ \\
  1.0$\times 10^0$ &  250 & 1.2$\times10^4$ &
  3.5$\times 10^4$ & 6.9$\times 10^{3}$ & 
  3.7$\times 10^3$ & 1.4$\times 10^3$ & 1.3$\times 10^3$ \\
  1.0$\times 10^0$ &  500 & 1.1$\times10^4$ &
  1.9$\times 10^4$ & 4.4$\times 10^{3}$ & 
  1.7$\times 10^3$ & 5.4$\times 10^2$ & 5.1$\times 10^2$ \\
  1.0$\times 10^0$ & 1000 & 1.1$\times10^4$ &
  9.8$\times 10^3$ & 2.3$\times 10^{3}$ & 
  1.1$\times 10^3$ & 2.8$\times 10^2$ & 2.8$\times 10^2$ \\
  1.0$\times 10^0$ & 3000 & 1.1$\times10^4$ &
  8.8$\times 10^3$ & 8.9$\times 10^{2}$ & 
  8.1$\times 10^2$ & 1.9$\times 10^2$ & 1.9$\times 10^2$ \\ \hline
  1.0$\times 10^1$ &   50 & 7.3$\times10^2$ &
  3.1$\times 10^3$ & 7.8$\times 10^{2}$ & 
  4.8$\times 10^2$ & 2.1$\times 10^2$ & 1.9$\times 10^2$ \\
  1.0$\times 10^1$ &  100 & 2.0$\times10^2$ &
  1.0$\times 10^3$ & 2.0$\times 10^{2}$ & 
  1.4$\times 10^2$ & 6.3$\times 10^1$ & 5.6$\times 10^1$ \\
  1.0$\times 10^1$ &  250 & 1.3$\times10^2$ &
  3.5$\times 10^2$ & 7.1$\times 10^{1}$ & 
  3.8$\times 10^1$ & 1.6$\times 10^1$ & 1.5$\times 10^1$ \\
  1.0$\times 10^1$ &  500 & 1.1$\times10^2$ &
  1.8$\times 10^2$ & 4.5$\times 10^{1}$ & 
  1.8$\times 10^1$ & 7.4$\times 10^0$ & 7.2$\times 10^0$ \\
  1.0$\times 10^1$ & 1000 & 1.1$\times10^2$ &
  1.0$\times 10^2$ & 2.5$\times 10^{1}$ & 
  1.2$\times 10^1$ & 4.9$\times 10^0$ & 4.9$\times 10^0$ \\
  1.0$\times 10^1$ & 3000 & 1.1$\times10^2$ &
  8.9$\times 10^1$ & 1.0$\times 10^{1}$ & 
  9.8$\times 10^0$ & 4.1$\times 10^0$ & 4.1$\times 10^0$ \\ \hline
  1.0$\times 10^2$ &   50 & 7.5$\times10^0$ &
  3.1$\times 10^1$ & 8.0$\times 10^{0}$ & 
  4.9$\times 10^0$ & 2.3$\times 10^0$ & 2.1$\times 10^0$ \\
  1.0$\times 10^2$ &  100 & 2.1$\times10^0$ &
  1.0$\times 10^1$ & 2.1$\times 10^{0}$ & 
  1.5$\times 10^0$ & 8.0$\times 10^{-1}$ & 7.3$\times 10^{-1}$ \\
  1.0$\times 10^2$ &  250 & 1.3$\times10^0$ &
  3.8$\times 10^0$ & 8.6$\times 10^{-1}$ & 
  5.1$\times 10^{-1}$ & 3.1$\times 10^{-1}$ & 3.0$\times 10^{-1}$ \\
  1.0$\times 10^2$ &  500 & 1.2$\times10^0$ &
  2.1$\times 10^0$ & 6.0$\times 10^{-1}$ & 
  3.2$\times 10^{-1}$ & 2.2$\times 10^{-1}$ & 2.1$\times 10^{-1}$ \\
  1.0$\times 10^2$ & 1000 & 1.2$\times10^0$ &
  1.1$\times 10^0$ & 3.8$\times 10^{-1}$ & 
  2.6$\times 10^{-1}$ & 1.8$\times 10^{-1}$ & 1.8$\times 10^{-1}$ \\
  1.0$\times 10^2$ & 3000 & 1.2$\times10^0$ & 
  1.0$\times 10^{0}$ & 2.3$\times 10^{-1}$ & 
  2.3$\times 10^{-1}$ & 1.7$\times 10^{-1}$ & 1.7$\times 10^{-1}$ \\ \hline
  1.0$\times 10^3$ &   50 & 9.7$\times10^{-2}$ &
  3.3$\times 10^{-1}$ & 9.5$\times 10^{-2}$ & 
  6.2$\times 10^{-2}$ & 3.8$\times 10^{-2}$ & 3.6$\times 10^{-2}$ \\
  1.0$\times 10^3$ &  100 & 3.2$\times10^{-2}$ &
  1.3$\times 10^{-1}$ & 3.4$\times 10^{-2}$ & 
  2.7$\times 10^{-2}$ & 2.1$\times 10^{-2}$ & 2.0$\times 10^{-2}$ \\
  1.0$\times 10^3$ &  250 & 2.2$\times10^{-2}$ &
  5.4$\times 10^{-2}$ & 2.0$\times 10^{-2}$ & 
  1.6$\times 10^{-2}$ & 1.4$\times 10^{-2}$ & 1.4$\times 10^{-2}$ \\
  1.0$\times 10^3$ &  500 & 2.1$\times10^{-2}$ &
  3.6$\times 10^{-2}$ & 1.7$\times 10^{-2}$ & 
  1.4$\times 10^{-2}$ & 1.2$\times 10^{-2}$ & 1.2$\times 10^{-2}$ \\
  1.0$\times 10^3$ & 1000 & 2.0$\times10^{-2}$ &
  2.3$\times 10^{-2}$ & 1.5$\times 10^{-2}$ & 
  1.3$\times 10^{-2}$ & 1.2$\times 10^{-2}$ & 1.2$\times 10^{-2}$ \\
  1.0$\times 10^3$ & 3000 & 2.0$\times10^{-2}$ &
  2.0$\times 10^{-2}$ & 1.3$\times 10^{-2}$ & 
  1.2$\times 10^{-2}$ & 1.1$\times 10^{-2}$ & 1.1$\times 10^{-2}$ \\ \hline
  1.0$\times 10^4$ &   50 & 3.1$\times10^{-3}$ &
  5.0$\times 10^{-3}$ & 2.2$\times 10^{-3}$ & 
  1.7$\times 10^{-3}$ & 1.5$\times 10^{-3}$ & 1.4$\times 10^{-3}$ \\
  1.0$\times 10^4$ &  100 & 1.5$\times10^{-3}$ &
  2.6$\times 10^{-3}$ & 1.4$\times 10^{-3}$ & 
  1.3$\times 10^{-3}$ & 1.2$\times 10^{-3}$ & 1.1$\times 10^{-3}$ \\
  1.0$\times 10^4$ &  250 & 1.1$\times10^{-3}$ &
  1.6$\times 10^{-3}$ & 1.1$\times 10^{-3}$ & 
  1.1$\times 10^{-3}$ & 1.1$\times 10^{-3}$ & 1.0$\times 10^{-3}$ \\
  1.0$\times 10^4$ &  500 & 1.1$\times10^{-3}$ &
  1.3$\times 10^{-3}$ & 1.1$\times 10^{-3}$ & 
  1.0$\times 10^{-3}$ & 9.9$\times 10^{-4}$ & 9.9$\times 10^{-4}$ \\
  1.0$\times 10^4$ & 1000 & 1.0$\times10^{-3}$ &
  1.1$\times 10^{-3}$ & 1.0$\times 10^{-3}$ & 
  9.9$\times 10^{-4}$ & 9.8$\times 10^{-4}$ & 9.8$\times 10^{-4}$ \\
  1.0$\times 10^4$ & 3000 & 1.0$\times10^{-3}$ &
  1.1$\times 10^{-3}$ & 1.0$\times 10^{-3}$ & 
  9.9$\times 10^{-4}$ & 9.7$\times 10^{-4}$ & 9.7$\times 10^{-4}$ \\ \hline
  1.0$\times 10^5$ &   50 & 2.5$\times10^{-4}$ &
  1.7$\times 10^{-4}$ & 1.3$\times 10^{-4}$ & 
  1.1$\times 10^{-4}$ & 1.1$\times 10^{-4}$ & 1.1$\times 10^{-4}$ \\
  1.0$\times 10^5$ &  100 & 1.3$\times10^{-4}$ &
  1.2$\times 10^{-4}$ & 1.0$\times 10^{-4}$ & 
  1.0$\times 10^{-4}$ & 9.9$\times 10^{-5}$ & 9.9$\times 10^{-5}$ \\
  1.0$\times 10^5$ &  250 & 1.0$\times10^{-4}$ &
  1.0$\times 10^{-4}$ & 9.6$\times 10^{-5}$ & 
  9.5$\times 10^{-5}$ & 9.6$\times 10^{-5}$ & 9.6$\times 10^{-5}$ \\
  1.0$\times 10^5$ &  500 & 9.6$\times10^{-5}$ &
  9.8$\times 10^{-5}$ & 9.4$\times 10^{-5}$ & 
  9.3$\times 10^{-5}$ & 9.3$\times 10^{-5}$ & 9.3$\times 10^{-5}$ \\
  1.0$\times 10^5$ & 1000 & 9.4$\times10^{-5}$ &
  9.5$\times 10^{-5}$ & 9.3$\times 10^{-5}$ & 
  9.3$\times 10^{-5}$ & 9.2$\times 10^{-5}$ & 9.2$\times 10^{-5}$ \\
  1.0$\times 10^5$ & 3000 & 9.4$\times10^{-5}$ &
  9.4$\times 10^{-5}$ & 9.3$\times 10^{-5}$ & 
  9.2$\times 10^{-5}$ & 9.2$\times 10^{-5}$ & 9.2$\times 10^{-5}$ \\ 
\end{tabular}
\caption{The minimal exposures needed to make a $3\sigma$ discovery of
  WIMP annihilation in the Sun when the atmospheric background
  is vertical. The values given are for hard muon spectra and a muon
  energy threshold of 1 GeV\@.}
  \label{tab:suver}
\end{table}

\begin{table}[h]
\begin{tabular}{lrllllll}
  \multicolumn{1}{l}{$\phi_s^0$} & 
  \multicolumn{1}{l}{$m_{\chi}$} & 
  \multicolumn{1}{l}{$\cE_{\rm min}^{\rm 0D}$} & 
  \multicolumn{1}{l}{$\cE_{\rm min}^{\rm 1D, 4 par}$} & 
  \multicolumn{1}{l}{$\cE_{\rm min}^{\rm 1D, 3 par}$} & 
  \multicolumn{1}{l}{$\cE_{\rm min}^{\rm 1D, 1 par}$} &
  \multicolumn{1}{l}{$\cE_{\rm min}^{\rm 2D, 3 par}$} &
  \multicolumn{1}{l}{$\cE_{\rm min}^{\rm 2D, 1 par}$} \\
  \multicolumn{1}{l}{[km$^{-2}$ yr$^{-1}$]} &
  \multicolumn{1}{l}{[GeV]} &
  \multicolumn{1}{l}{[km$^{2}$ yr]} &
  \multicolumn{1}{l}{[km$^{2}$ yr]} &
  \multicolumn{1}{l}{[km$^{2}$ yr]} &
  \multicolumn{1}{l}{[km$^{2}$ yr]} &
  \multicolumn{1}{l}{[km$^{2}$ yr]} &
  \multicolumn{1}{l}{[km$^{2}$ yr]} \\ \hline
  1.0$\times 10^0$ &   50 & 1.7$\times10^5$ &
  7.2$\times 10^5$ & 1.8$\times 10^{5}$ & 
  1.1$\times 10^5$ & 4.7$\times 10^4$ & 4.3$\times 10^4$ \\
  1.0$\times 10^0$ &  100 & 4.6$\times10^4$ &
  2.3$\times 10^5$ & 4.5$\times 10^{4}$ & 
  3.2$\times 10^4$ & 1.5$\times 10^4$ & 1.3$\times 10^4$ \\
  1.0$\times 10^0$ &  250 & 2.9$\times10^4$ &
  8.0$\times 10^4$ & 1.6$\times 10^{4}$ & 
  8.4$\times 10^3$ & 3.8$\times 10^3$ & 3.4$\times 10^3$ \\
  1.0$\times 10^0$ &  500 & 2.6$\times10^4$ &
  4.2$\times 10^4$ & 1.0$\times 10^{4}$ & 
  3.8$\times 10^3$ & 1.4$\times 10^3$ & 1.4$\times 10^3$ \\
  1.0$\times 10^0$ & 1000 & 2.5$\times10^4$ &
  2.3$\times 10^4$ & 5.3$\times 10^{3}$ & 
  2.5$\times 10^3$ & 7.7$\times 10^2$ & 7.7$\times 10^2$ \\
  1.0$\times 10^0$ & 3000 & 2.5$\times10^4$ &
  2.0$\times 10^4$ & 2.0$\times 10^{3}$ & 
  1.8$\times 10^3$ & 5.3$\times 10^2$ & 5.2$\times 10^2$ \\ \hline
  1.0$\times 10^1$ &   50 & 1.7$\times10^3$ &
  7.2$\times 10^3$ & 1.8$\times 10^{3}$ & 
  1.1$\times 10^3$ & 4.7$\times 10^2$ & 4.3$\times 10^2$ \\
  1.0$\times 10^1$ &  100 & 4.6$\times10^2$ &
  2.4$\times 10^3$ & 4.5$\times 10^{2}$ & 
  3.2$\times 10^2$ & 1.5$\times 10^2$ & 1.3$\times 10^2$ \\
  1.0$\times 10^1$ &  250 & 2.9$\times10^2$ &
  8.0$\times 10^2$ & 1.6$\times 10^{2}$ & 
  8.6$\times 10^1$ & 4.0$\times 10^1$ & 3.6$\times 10^1$ \\
  1.0$\times 10^1$ &  500 & 2.6$\times10^2$ &
  4.2$\times 10^2$ & 1.0$\times 10^{2}$ & 
  4.0$\times 10^1$ & 1.7$\times 10^1$ & 1.6$\times 10^1$ \\
  1.0$\times 10^1$ & 1000 & 2.5$\times10^2$ &
  2.2$\times 10^2$ & 5.5$\times 10^{1}$ & 
  2.7$\times 10^1$ & 1.0$\times 10^1$ & 1.0$\times 10^1$ \\
  1.0$\times 10^1$ & 3000 & 2.5$\times10^2$ &
  2.0$\times 10^2$ & 2.2$\times 10^{1}$ & 
  2.0$\times 10^1$ & 7.7$\times 10^0$ & 7.7$\times 10^0$ \\ \hline
  1.0$\times 10^2$ &   50 & 1.7$\times10^1$ &
  7.2$\times 10^1$ & 1.8$\times 10^{1}$ & 
  1.1$\times 10^1$ & 4.9$\times 10^0$ & 4.5$\times 10^0$ \\
  1.0$\times 10^2$ &  100 & 4.7$\times10^0$ &
  2.4$\times 10^1$ & 4.7$\times 10^{0}$ & 
  3.3$\times 10^0$ & 1.6$\times 10^{0}$ & 1.5$\times 10^{0}$ \\
  1.0$\times 10^2$ &  250 & 3.0$\times10^0$ &
  8.3$\times 10^0$ & 1.8$\times 10^{0}$ & 
  1.0$\times 10^{0}$ & 5.6$\times 10^{-1}$ & 5.2$\times 10^{-1}$ \\
  1.0$\times 10^2$ &  500 & 2.7$\times10^0$ &
  4.5$\times 10^0$ & 1.2$\times 10^{0}$ & 
  5.4$\times 10^{-1}$ & 3.3$\times 10^{-1}$ & 3.2$\times 10^{-1}$ \\
  1.0$\times 10^2$ & 1000 & 2.6$\times10^0$ &
  2.4$\times 10^0$ & 6.9$\times 10^{-1}$ & 
  4.1$\times 10^{-1}$ & 2.6$\times 10^{-1}$ & 2.6$\times 10^{-1}$ \\
  1.0$\times 10^2$ & 3000 & 2.6$\times10^0$ &
  2.1$\times 10^{0}$ & 3.5$\times 10^{-1}$ & 
  3.5$\times 10^{-1}$ & 2.4$\times 10^{-1}$ & 2.4$\times 10^{-1}$ \\ \hline
  1.0$\times 10^3$ &   50 & 1.9$\times10^{-1}$ &
  7.4$\times 10^{-1}$ & 2.0$\times 10^{-1}$ & 
  1.2$\times 10^{-1}$ & 6.5$\times 10^{-2}$ & 6.0$\times 10^{-2}$ \\
  1.0$\times 10^3$ &  100 & 5.8$\times10^{-2}$ &
  2.6$\times 10^{-1}$ & 6.0$\times 10^{-2}$ & 
  4.5$\times 10^{-2}$ & 3.1$\times 10^{-2}$ & 2.9$\times 10^{-2}$ \\
  1.0$\times 10^3$ &  250 & 3.8$\times10^{-2}$ &
  1.0$\times 10^{-1}$ & 3.0$\times 10^{-2}$ & 
  2.2$\times 10^{-2}$ & 1.8$\times 10^{-2}$ & 1.7$\times 10^{-2}$ \\
  1.0$\times 10^3$ &  500 & 3.5$\times10^{-2}$ &
  6.4$\times 10^{-2}$ & 2.4$\times 10^{-2}$ & 
  1.7$\times 10^{-2}$ & 1.5$\times 10^{-2}$ & 1.4$\times 10^{-2}$ \\
  1.0$\times 10^3$ & 1000 & 3.5$\times10^{-2}$ &
  3.8$\times 10^{-2}$ & 1.9$\times 10^{-2}$ & 
  1.5$\times 10^{-2}$ & 1.3$\times 10^{-2}$ & 1.3$\times 10^{-2}$ \\
  1.0$\times 10^3$ & 3000 & 3.4$\times10^{-2}$ &
  3.2$\times 10^{-2}$ & 1.5$\times 10^{-2}$ & 
  1.5$\times 10^{-2}$ & 1.3$\times 10^{-2}$ & 1.3$\times 10^{-2}$ \\ \hline
  1.0$\times 10^4$ &   50 & 4.1$\times10^{-3}$ &
  9.3$\times 10^{-3}$ & 3.4$\times 10^{-3}$ & 
  2.4$\times 10^{-3}$ & 1.9$\times 10^{-3}$ & 1.8$\times 10^{-3}$ \\
  1.0$\times 10^4$ &  100 & 1.7$\times10^{-3}$ &
  4.2$\times 10^{-3}$ & 1.7$\times 10^{-3}$ & 
  1.5$\times 10^{-3}$ & 1.3$\times 10^{-3}$ & 1.3$\times 10^{-3}$ \\
  1.0$\times 10^4$ &  250 & 1.3$\times10^{-3}$ &
  2.2$\times 10^{-3}$ & 1.3$\times 10^{-3}$ & 
  1.2$\times 10^{-3}$ & 1.1$\times 10^{-3}$ & 1.1$\times 10^{-3}$ \\
  1.0$\times 10^4$ &  500 & 1.2$\times10^{-3}$ &
  1.7$\times 10^{-3}$ & 1.2$\times 10^{-3}$ & 
  1.1$\times 10^{-3}$ & 1.1$\times 10^{-3}$ & 1.0$\times 10^{-3}$ \\
  1.0$\times 10^4$ & 1000 & 1.2$\times10^{-3}$ &
  1.4$\times 10^{-3}$ & 1.1$\times 10^{-3}$ & 
  1.0$\times 10^{-3}$ & 1.0$\times 10^{-3}$ & 1.0$\times 10^{-3}$ \\
  1.0$\times 10^4$ & 3000 & 1.2$\times10^{-3}$ &
  1.2$\times 10^{-3}$ & 1.1$\times 10^{-3}$ & 
  1.0$\times 10^{-3}$ & 1.0$\times 10^{-3}$ & 1.0$\times 10^{-3}$ \\ \hline
  1.0$\times 10^5$ &   50 & 2.6$\times10^{-4}$ &
  2.3$\times 10^{-4}$ & 1.5$\times 10^{-4}$ & 
  1.2$\times 10^{-4}$ & 1.2$\times 10^{-4}$ & 1.1$\times 10^{-4}$ \\
  1.0$\times 10^5$ &  100 & 1.3$\times10^{-4}$ &
  1.5$\times 10^{-4}$ & 1.1$\times 10^{-4}$ & 
  1.1$\times 10^{-4}$ & 1.0$\times 10^{-4}$ & 1.0$\times 10^{-4}$ \\
  1.0$\times 10^5$ &  250 & 1.0$\times10^{-4}$ &
  1.1$\times 10^{-4}$ & 9.9$\times 10^{-5}$ & 
  9.8$\times 10^{-5}$ & 9.8$\times 10^{-5}$ & 9.8$\times 10^{-5}$ \\
  1.0$\times 10^5$ &  500 & 9.7$\times10^{-5}$ &
  1.0$\times 10^{-4}$ & 9.6$\times 10^{-5}$ & 
  9.5$\times 10^{-5}$ & 9.5$\times 10^{-5}$ & 9.4$\times 10^{-5}$ \\
  1.0$\times 10^5$ & 1000 & 9.6$\times10^{-5}$ &
  9.9$\times 10^{-5}$ & 9.5$\times 10^{-5}$ & 
  9.4$\times 10^{-5}$ & 9.4$\times 10^{-5}$ & 9.4$\times 10^{-5}$ \\
  1.0$\times 10^5$ & 3000 & 9.5$\times10^{-5}$ &
  9.7$\times 10^{-5}$ & 9.4$\times 10^{-5}$ & 
  9.4$\times 10^{-5}$ & 9.3$\times 10^{-5}$ & 9.3$\times 10^{-5}$ \\ 
\end{tabular}
\caption{The minimal exposures needed to make a $3\sigma$ discovery of
  WIMP annihilation in the Sun when the atmospheric background
  is horizontal. The values given are for hard muon spectra and a muon
  energy threshold of 1 GeV\@.}
  \label{tab:suhor}
\end{table}

\begin{table}[h]
\begin{tabular}{lrllllll}
  \multicolumn{1}{l}{$\phi_s^0$} & 
  \multicolumn{1}{l}{$m_{\chi}$} & 
  \multicolumn{1}{l}{$\cE_{\rm min}^{\rm 0D}$} & 
  \multicolumn{1}{l}{$\cE_{\rm min}^{\rm 1D, 4 par}$} & 
  \multicolumn{1}{l}{$\cE_{\rm min}^{\rm 1D, 3 par}$} & 
  \multicolumn{1}{l}{$\cE_{\rm min}^{\rm 1D, 1 par}$} &
  \multicolumn{1}{l}{$\cE_{\rm min}^{\rm 2D, 3 par}$} &
  \multicolumn{1}{l}{$\cE_{\rm min}^{\rm 2D, 1 par}$} \\
  \multicolumn{1}{l}{[km$^{-2}$ yr$^{-1}$]} &
  \multicolumn{1}{l}{[GeV]} &
  \multicolumn{1}{l}{[km$^{2}$ yr]} &
  \multicolumn{1}{l}{[km$^{2}$ yr]} &
  \multicolumn{1}{l}{[km$^{2}$ yr]} &
  \multicolumn{1}{l}{[km$^{2}$ yr]} &
  \multicolumn{1}{l}{[km$^{2}$ yr]} &
  \multicolumn{1}{l}{[km$^{2}$ yr]} \\ \hline
  1.0$\times 10^0$ &   50 & 1.4$\times10^5$ &
  5.2$\times 10^5$ & 1.9$\times 10^{5}$ & 
  7.2$\times 10^4$ & 3.9$\times 10^4$ & 3.4$\times 10^4$ \\
  1.0$\times 10^0$ &  100 & 3.1$\times10^4$ &
  2.5$\times 10^5$ & 5.2$\times 10^{4}$ & 
  2.4$\times 10^4$ & 1.3$\times 10^4$ & 1.1$\times 10^4$ \\
  1.0$\times 10^0$ &  250 & 1.3$\times10^4$ &
  5.7$\times 10^4$ & 1.6$\times 10^{4}$ & 
  7.6$\times 10^3$ & 3.3$\times 10^3$ & 3.0$\times 10^3$ \\
  1.0$\times 10^0$ &  500 & 1.1$\times10^4$ &
  2.5$\times 10^4$ & 7.4$\times 10^{3}$ & 
  3.5$\times 10^3$ & 1.4$\times 10^3$ & 1.2$\times 10^3$ \\
  1.0$\times 10^0$ & 1000 & 1.1$\times10^4$ &
  1.4$\times 10^4$ & 3.9$\times 10^{3}$ & 
  1.7$\times 10^3$ & 4.7$\times 10^2$ & 3.9$\times 10^2$ \\
  1.0$\times 10^0$ & 3000 & 1.0$\times10^4$ &
  4.8$\times 10^3$ & 1.3$\times 10^{3}$ & 
  5.2$\times 10^2$ & 6.8$\times 10^2$ & 6.2$\times 10^1$ \\ \hline
  1.0$\times 10^1$ &   50 & 1.4$\times10^3$ &
  5.2$\times 10^3$ & 1.9$\times 10^{3}$ & 
  7.3$\times 10^2$ & 3.9$\times 10^2$ & 3.4$\times 10^2$ \\
  1.0$\times 10^1$ &  100 & 3.1$\times10^2$ &
  2.5$\times 10^3$ & 5.3$\times 10^{2}$ & 
  2.4$\times 10^2$ & 1.3$\times 10^2$ & 1.1$\times 10^2$ \\
  1.0$\times 10^1$ &  250 & 1.3$\times10^2$ &
  5.8$\times 10^2$ & 1.6$\times 10^{2}$ & 
  7.7$\times 10^1$ & 3.5$\times 10^1$ & 3.1$\times 10^1$ \\
  1.0$\times 10^1$ &  500 & 1.1$\times10^2$ &
  2.5$\times 10^2$ & 7.5$\times 10^{1}$ & 
  3.7$\times 10^1$ & 1.5$\times 10^1$ & 1.3$\times 10^1$  \\
  1.0$\times 10^1$ & 1000 & 1.1$\times10^2$ &
  1.4$\times 10^2$ & 4.0$\times 10^{1}$ & 
  1.8$\times 10^1$ & 6.5$\times 10^0$ & 5.6$\times 10^0$ \\
  1.0$\times 10^1$ & 3000 & 1.0$\times10^2$ &
  4.9$\times 10^1$ & 1.4$\times 10^{1}$ & 
  6.5$\times 10^0$ & 2.3$\times 10^0$ & 2.1$\times 10^0$ \\ \hline
  1.0$\times 10^2$ &   50 & 1.4$\times10^1$ &
  5.3$\times 10^1$ & 2.0$\times 10^{1}$ & 
  7.4$\times 10^0$ & 4.1$\times 10^0$ & 3.5$\times 10^0$ \\
  1.0$\times 10^2$ &  100 & 3.2$\times10^0$ &
  2.5$\times 10^1$ & 5.4$\times 10^{0}$ & 
  2.6$\times 10^0$ & 1.5$\times 10^{0}$ & 1.3$\times 10^{0}$ \\
  1.0$\times 10^2$ &  250 & 1.4$\times10^0$ &
  6.0$\times 10^0$ & 1.8$\times 10^{0}$ & 
  8.9$\times 10^{-1}$ & 4.9$\times 10^{-1}$ & 4.5$\times 10^{-1}$ \\
  1.0$\times 10^2$ &  500 & 1.2$\times10^0$ &
  2.7$\times 10^0$ & 8.7$\times 10^{-1}$ & 
  4.8$\times 10^{-1}$ & 2.8$\times 10^{-1}$ & 2.6$\times 10^{-1}$ \\
  1.0$\times 10^2$ & 1000 & 1.1$\times10^0$ &
  1.5$\times 10^0$ & 5.0$\times 10^{-1}$ & 
  3.0$\times 10^{-1}$ & 1.9$\times 10^{-1}$ & 1.8$\times 10^{-1}$ \\
  1.0$\times 10^2$ & 3000 & 1.1$\times10^0$ &
  6.0$\times 10^{-1}$ & 2.4$\times 10^{-1}$ & 
  1.7$\times 10^{-1}$ & 1.2$\times 10^{-1}$ & 1.2$\times 10^{-1}$ \\ \hline
  1.0$\times 10^3$ &   50 & 1.7$\times10^{-1}$ &
  5.6$\times 10^{-1}$ & 2.1$\times 10^{-1}$ & 
  8.6$\times 10^{-2}$ & 5.7$\times 10^{-2}$ & 5.0$\times 10^{-2}$ \\
  1.0$\times 10^3$ &  100 & 4.7$\times10^{-2}$ &
  2.8$\times 10^{-1}$ & 6.6$\times 10^{-2}$ & 
  3.7$\times 10^{-2}$ & 2.7$\times 10^{-2}$ & 2.5$\times 10^{-2}$ \\
  1.0$\times 10^3$ &  250 & 2.3$\times10^{-2}$ &
  7.5$\times 10^{-2}$ & 2.8$\times 10^{-2}$ & 
  1.9$\times 10^{-2}$ & 1.6$\times 10^{-2}$ & 1.5$\times 10^{-2}$ \\
  1.0$\times 10^3$ &  500 & 2.0$\times10^{-2}$ &
  3.9$\times 10^{-2}$ & 1.9$\times 10^{-2}$ & 
  1.5$\times 10^{-2}$ & 1.3$\times 10^{-2}$ & 1.3$\times 10^{-2}$ \\
  1.0$\times 10^3$ & 1000 & 2.0$\times10^{-2}$ &
  2.5$\times 10^{-2}$ & 1.4$\times 10^{-2}$ & 
  1.3$\times 10^{-2}$ & 1.1$\times 10^{-2}$ & 1.1$\times 10^{-2}$ \\
  1.0$\times 10^3$ & 3000 & 1.9$\times10^{-2}$ &
  1.5$\times 10^{-2}$ & 1.1$\times 10^{-2}$ & 
  1.1$\times 10^{-2}$ & 9.9$\times 10^{-3}$ & 9.9$\times 10^{-3}$ \\ \hline
  1.0$\times 10^4$ &   50 & 4.7$\times10^{-3}$ &
  8.2$\times 10^{-3}$ & 3.5$\times 10^{-3}$ & 
  1.9$\times 10^{-3}$ & 1.7$\times 10^{-3}$ & 1.6$\times 10^{-3}$ \\
  1.0$\times 10^4$ &  100 & 1.9$\times10^{-3}$ &
  4.3$\times 10^{-3}$ & 1.7$\times 10^{-3}$ & 
  1.4$\times 10^{-3}$ & 1.3$\times 10^{-3}$ & 1.2$\times 10^{-3}$ \\
  1.0$\times 10^4$ &  250 & 1.2$\times10^{-3}$ &
  1.8$\times 10^{-3}$ & 1.2$\times 10^{-3}$ & 
  1.1$\times 10^{-3}$ & 1.1$\times 10^{-3}$ & 1.1$\times 10^{-3}$ \\
  1.0$\times 10^4$ &  500 & 1.0$\times10^{-3}$ &
  1.3$\times 10^{-3}$ & 1.1$\times 10^{-3}$ & 
  1.0$\times 10^{-3}$ & 1.0$\times 10^{-3}$ & 1.0$\times 10^{-3}$ \\
  1.0$\times 10^4$ & 1000 & 1.0$\times10^{-3}$ &
  1.1$\times 10^{-3}$ & 9.9$\times 10^{-4}$ & 
  9.7$\times 10^{-4}$ & 9.6$\times 10^{-4}$ & 9.6$\times 10^{-4}$ \\
  1.0$\times 10^4$ & 3000 & 1.0$\times10^{-3}$ &
  9.8$\times 10^{-4}$ & 9.4$\times 10^{-4}$ & 
  9.3$\times 10^{-4}$ & 9.2$\times 10^{-4}$ & 9.2$\times 10^{-4}$ \\ \hline
  1.0$\times 10^5$ &   50 & 3.5$\times10^{-4}$ &
  2.3$\times 10^{-4}$ & 1.5$\times 10^{-4}$ & 
  1.1$\times 10^{-4}$ & 1.1$\times 10^{-4}$ & 1.1$\times 10^{-4}$ \\
  1.0$\times 10^5$ &  100 & 1.6$\times10^{-4}$ &
  1.4$\times 10^{-4}$ & 1.1$\times 10^{-4}$ & 
  1.0$\times 10^{-4}$ & 1.0$\times 10^{-4}$ & 9.9$\times 10^{-5}$ \\
  1.0$\times 10^5$ &  250 & 1.0$\times10^{-4}$ &
  1.0$\times 10^{-4}$ & 9.6$\times 10^{-5}$ & 
  9.5$\times 10^{-5}$ & 9.5$\times 10^{-5}$ & 9.4$\times 10^{-5}$ \\
  1.0$\times 10^5$ &  500 & 9.5$\times10^{-5}$ &
  9.7$\times 10^{-5}$ & 9.3$\times 10^{-5}$ & 
  9.3$\times 10^{-5}$ & 9.3$\times 10^{-5}$ & 9.3$\times 10^{-5}$ \\
  1.0$\times 10^5$ & 1000 & 9.2$\times10^{-5}$ &
  9.3$\times 10^{-5}$ & 9.2$\times 10^{-5}$ & 
  9.2$\times 10^{-5}$ & 9.2$\times 10^{-5}$ & 9.2$\times 10^{-5}$ \\
  1.0$\times 10^5$ & 3000 & 9.1$\times10^{-5}$ &
  9.1$\times 10^{-5}$ & 9.1$\times 10^{-5}$ & 
  9.0$\times 10^{-5}$ & 9.0$\times 10^{-5}$ & 9.0$\times 10^{-5}$ \\ 
\end{tabular}
\caption{The minimal exposures needed to make a $3\sigma$ discovery of
  WIMP annihilation in the Earth with the atmospheric background
  being vertical. The values given are for hard muon spectra and a muon
  energy threshold of 1 GeV\@.}
  \label{tab:eaver}
\end{table}

\end{document}